\pdfoutput=1
\documentclass[12pt,a4paper]{article}

\usepackage{booktabs}
\usepackage{ifthen} 
\usepackage{makecell}
\newboolean{pdflatex}
\setboolean{pdflatex}{true} 

\newboolean{articletitles}
\setboolean{articletitles}{true} 

\newboolean{uprightparticles}
\setboolean{uprightparticles}{false} 

\def\paperauthors{LHCb collaboration} 
\def\paperasciititle{Measurement of CP violation in the decay B to K+ pi0} 
\def\papertitle{Measurement of \CP violation \\ in the decay \sig} 
\def\paperkeywords{{High Energy Physics}, {LHCb}} 
\def\papercopyright{\the\year\ CERN for the benefit of the LHCb collaboration} 
\def\paperlicence{CC BY 4.0 licence}
\def\paperlicenceurl{https://creativecommons.org/licenses/by/4.0/}

\usepackage[top=1in, bottom=1.25in, left=1in, right=1in]{geometry}

\usepackage{microtype}
\usepackage{lineno}  
\usepackage{xspace} 
\usepackage{caption} 

\usepackage{graphicx}  
\usepackage{color}
\usepackage{colortbl}
\graphicspath{{./figs/}} 

\usepackage{amsmath} 
\usepackage{amssymb}
\usepackage{amsfonts}
\usepackage{upgreek} 

\newcommand*\patchAmsMathEnvironmentForLineno[1]{%
\expandafter\let\csname old#1\expandafter\endcsname\csname #1\endcsname
\expandafter\let\csname oldend#1\expandafter\endcsname\csname
end#1\endcsname
 \renewenvironment{#1}%
   {\linenomath\csname old#1\endcsname}%
   {\csname oldend#1\endcsname\endlinenomath}%
}
\newcommand*\patchBothAmsMathEnvironmentsForLineno[1]{%
  \patchAmsMathEnvironmentForLineno{#1}%
  \patchAmsMathEnvironmentForLineno{#1*}%
}
\AtBeginDocument{%
\patchBothAmsMathEnvironmentsForLineno{equation}%
\patchBothAmsMathEnvironmentsForLineno{align}%
\patchBothAmsMathEnvironmentsForLineno{flalign}%
\patchBothAmsMathEnvironmentsForLineno{alignat}%
\patchBothAmsMathEnvironmentsForLineno{gather}%
\patchBothAmsMathEnvironmentsForLineno{multline}%
\patchBothAmsMathEnvironmentsForLineno{eqnarray}%
}

\usepackage{hyperxmp}

\usepackage[pdftex,
            pdfauthor={\paperauthors},
            pdftitle={\paperasciititle},
            pdfkeywords={\paperkeywords},
            pdfcopyright={Copyright (C) \papercopyright},
            pdflicenseurl={\paperlicenceurl}]{hyperref}

\usepackage[colorinlistoftodos,textsize=scriptsize]{todonotes}

\usepackage[bottom,flushmargin,hang,multiple]{footmisc}

\usepackage[all]{hypcap} 

\usepackage{xspace} 
\usepackage{upgreek}

\def\lhcb   {\mbox{LHCb}\xspace}

\def\MagUp {\mbox{\em Mag\kern -0.05em Up}\xspace}

\ifthenelse{\boolean{uprightparticles}}%
{

 \def\Ppi         {\ensuremath{\uppi}\xspace}

 \def\Ppsi        {\ensuremath{\uppsi}\xspace}

 \def\PDelta      {\ensuremath{\Delta}\xspace}                 
 \def\PXi         {\ensuremath{\Xi}\xspace}                 
 \def\PLambda     {\ensuremath{\Lambda}\xspace}                 
 \def\PSigma      {\ensuremath{\Sigma}\xspace}                 
 \def\POmega      {\ensuremath{\Omega}\xspace}                 
 \def\PUpsilon    {\ensuremath{\Upsilon}\xspace}

 \def\PB      {\ensuremath{\mathrm{B}}\xspace}                 
                  
 \def\PD      {\ensuremath{\mathrm{D}}\xspace}

 \def\PJ      {\ensuremath{\mathrm{J}}\xspace}                 
 \def\PK      {\ensuremath{\mathrm{K}}\xspace}

 \def\Pb      {\ensuremath{\mathrm{b}}\xspace}

 \def\Pi      {\ensuremath{\mathrm{i}}\xspace}

 \def\Ps      {\ensuremath{\mathrm{s}}\xspace}

 \def\thebaroffset{0.0em}
}
{

 \def\Ppi         {\ensuremath{\pi}\xspace}

 \def\Ppsi        {\ensuremath{\psi}\xspace}                 
                  
 \mathchardef\PDelta="7101
 \mathchardef\PXi="7104
 \mathchardef\PLambda="7103
 \mathchardef\PSigma="7106
 \mathchardef\POmega="710A
 \mathchardef\PUpsilon="7107
                  
 \def\PB      {\ensuremath{B}\xspace}                 
                  
 \def\PD      {\ensuremath{D}\xspace}

 \def\PJ      {\ensuremath{J}\xspace}                 
 \def\PK      {\ensuremath{K}\xspace}

 \def\Pb      {\ensuremath{b}\xspace}

 \def\Pi      {\ensuremath{i}\xspace}

 \def\Ps      {\ensuremath{s}\xspace}

 \def\thebaroffset{0.18em}
}
\newcommand{\offsetoverline}[2][\thebaroffset]{\kern #1\overline{\kern -#1 #2}}%

\makeatletter
\ifcase \@ptsize \relax
  \newcommand{\miniscule}{\@setfontsize\miniscule{4}{5}}
\or
  \newcommand{\miniscule}{\@setfontsize\miniscule{5}{6}}
\or
  \newcommand{\miniscule}{\@setfontsize\miniscule{5}{6}}
\fi
\makeatother

\DeclareRobustCommand{\optbar}[1]{\shortstack{{\miniscule (\rule[.5ex]{1.25em}{.18mm})}
  \\ [-.7ex] $#1$}}

\def\squark    {{\ensuremath{\Ps}}\xspace}

\def\bquark    {{\ensuremath{\Pb}}\xspace}

\def\pion   {{\ensuremath{\Ppi}}\xspace}
\def\piz    {{\ensuremath{\pion^0}}\xspace}
\def\pip    {{\ensuremath{\pion^+}}\xspace}
\def\pim    {{\ensuremath{\pion^-}}\xspace}

\def\kaon    {{\ensuremath{\PK}}\xspace}

\def\KorKbar {\kern \thebaroffset\optbar{\kern -\thebaroffset \PK}{}\xspace}
\def\Kz      {{\ensuremath{\kaon^0}}\xspace}

\def\Kp      {{\ensuremath{\kaon^+}}\xspace}
\def\Km      {{\ensuremath{\kaon^-}}\xspace}

\def\D       {{\ensuremath{\PD}}\xspace}

\def\DorDbar {\kern \thebaroffset\optbar{\kern -\thebaroffset \PD}\xspace}

\def\Dp      {{\ensuremath{\D^+}}\xspace}
\def\Dm      {{\ensuremath{\D^-}}\xspace}

\def\DpDm    {\ensuremath{\Dp {\kern -0.16em \Dm}}\xspace}

\def\B       {{\ensuremath{\PB}}\xspace}

\def\BorBbar {\kern \thebaroffset\optbar{\kern -\thebaroffset \PB}\xspace}
\def\Bz      {{\ensuremath{\B^0}}\xspace}

\def\Bd      {{\ensuremath{\B^0}}\xspace}

\def\BdorBdbar {\kern \thebaroffset\optbar{\kern -\thebaroffset \Bd}\xspace}
\def\Bu      {{\ensuremath{\B^+}}\xspace}
\def\Bub     {{\ensuremath{\B^-}}\xspace}
\def\Bp      {{\ensuremath{\Bu}}\xspace}
\def\Bm      {{\ensuremath{\Bub}}\xspace}

\def\Bs      {{\ensuremath{\B^0_\squark}}\xspace}

\def\BsorBsbar {\kern \thebaroffset\optbar{\kern -\thebaroffset \Bs}\xspace}

\def\jpsi     {{\ensuremath{{\PJ\mskip -3mu/\mskip -2mu\Ppsi}}}\xspace}

\def\Y#1S{\ensuremath{\PUpsilon{(#1S)}}\xspace}

\def\LorLbar     {\kern \thebaroffset\optbar{\kern -\thebaroffset \PLambda}\xspace}

\def\BF         {{\ensuremath{\mathcal{B}}}\xspace}

\newcommand{\decay}[2]{\ensuremath{#1\!\to #2}\xspace} 

\def\to                 {\ensuremath{\rightarrow}\xspace}

\def\CP                {{\ensuremath{C\!P}}\xspace}

\def\BdToKpi      {\decay{\Bd}{\Kp\pim}}

\def\AT#1     {\ensuremath{A_{\mathrm{T}}^{#1}}\xspace}           

\def\C#1      {\ensuremath{\mathcal{C}_{#1}}\xspace}                       
\def\Cp#1     {\ensuremath{\mathcal{C}_{#1}^{'}}\xspace}                    
\def\Ceff#1   {\ensuremath{\mathcal{C}_{#1}^{\mathrm{(eff)}}}\xspace}        
\def\Cpeff#1  {\ensuremath{\mathcal{C}_{#1}^{'\mathrm{(eff)}}}\xspace}       
\def\Ope#1    {\ensuremath{\mathcal{O}_{#1}}\xspace}                       
\def\Opep#1   {\ensuremath{\mathcal{O}_{#1}^{'}}\xspace}                    

\newcommand{\nospaceunit}[1]{\ensuremath{\text{#1}}}       
\newcommand{\aunit}[1]{\ensuremath{\text{\,#1}}}       

\newcommand{\tev}{\aunit{Te\kern -0.1em V}\xspace}
\newcommand{\gev}{\aunit{Ge\kern -0.1em V}\xspace}
\newcommand{\mev}{\aunit{Me\kern -0.1em V}\xspace}
\newcommand{\kev}{\aunit{ke\kern -0.1em V}\xspace}
\newcommand{\ev}{\aunit{e\kern -0.1em V}\xspace}
 
\newcommand{\mevc}{\ensuremath{\aunit{Me\kern -0.1em V\!/}c}\xspace}
\newcommand{\gevc}{\ensuremath{\aunit{Ge\kern -0.1em V\!/}c}\xspace}
\newcommand{\mevcc}{\ensuremath{\aunit{Me\kern -0.1em V\!/}c^2}\xspace}
\newcommand{\gevcc}{\ensuremath{\aunit{Ge\kern -0.1em V\!/}c^2}\xspace}

\def\mum  {\ensuremath{\,\upmu\nospaceunit{m}}\xspace}

\def\fb   {\ensuremath{\aunit{fb}}\xspace}
\def\invfb   {\ensuremath{\fb^{-1}}\xspace}

\def\ps   {\ensuremath{\aunit{ps}}\xspace}

\newcommand{\chisq}{\ensuremath{\chi^2}\xspace}

\def\gsim{{~\raise.15em\hbox{$>$}\kern-.85em
          \lower.35em\hbox{$\sim$}~}\xspace}
\def\lsim{{~\raise.15em\hbox{$<$}\kern-.85em
          \lower.35em\hbox{$\sim$}~}\xspace}

\def\pt         {\ensuremath{p_{\mathrm{T}}}\xspace}

\def\ptot       {\ensuremath{p}\xspace}

\def\evtgen     {\mbox{\textsc{EvtGen}}\xspace}

\def\geant      {\mbox{\textsc{Geant4}}\xspace}

\def\photos     {\mbox{\textsc{Photos}}\xspace}

\def\pythia     {\mbox{\textsc{Pythia}}\xspace}

\def\tell1  {TELL1\xspace}
\def\ukl1   {UKL1\xspace}

\usepackage{cite} 
\usepackage{mciteplus}

\usepackage{longtable} 
\mciteErrorOnUnknownfalse
\usepackage{subcaption}
\newcommand{\sig}{\mbox{\decay{\Bu}{\Kp\piz}}}
\newcommand{\Apt}{\ensuremath{{\cal A}(\pt)}}
\usepackage{multirow}

\begin{document}

\renewcommand{\thefootnote}{\fnsymbol{footnote}}
\setcounter{footnote}{1}


\begin{titlepage}
\pagenumbering{roman}

\vspace*{-1.5cm}
\centerline{\large EUROPEAN ORGANIZATION FOR NUCLEAR RESEARCH (CERN)}
\vspace*{1.5cm}
\noindent
\begin{tabular*}{\linewidth}{lc@{\extracolsep{\fill}}r@{\extracolsep{0pt}}}
\ifthenelse{\boolean{pdflatex}}
{\vspace*{-1.5cm}\mbox{\!\!\!\includegraphics[width=.14\textwidth]{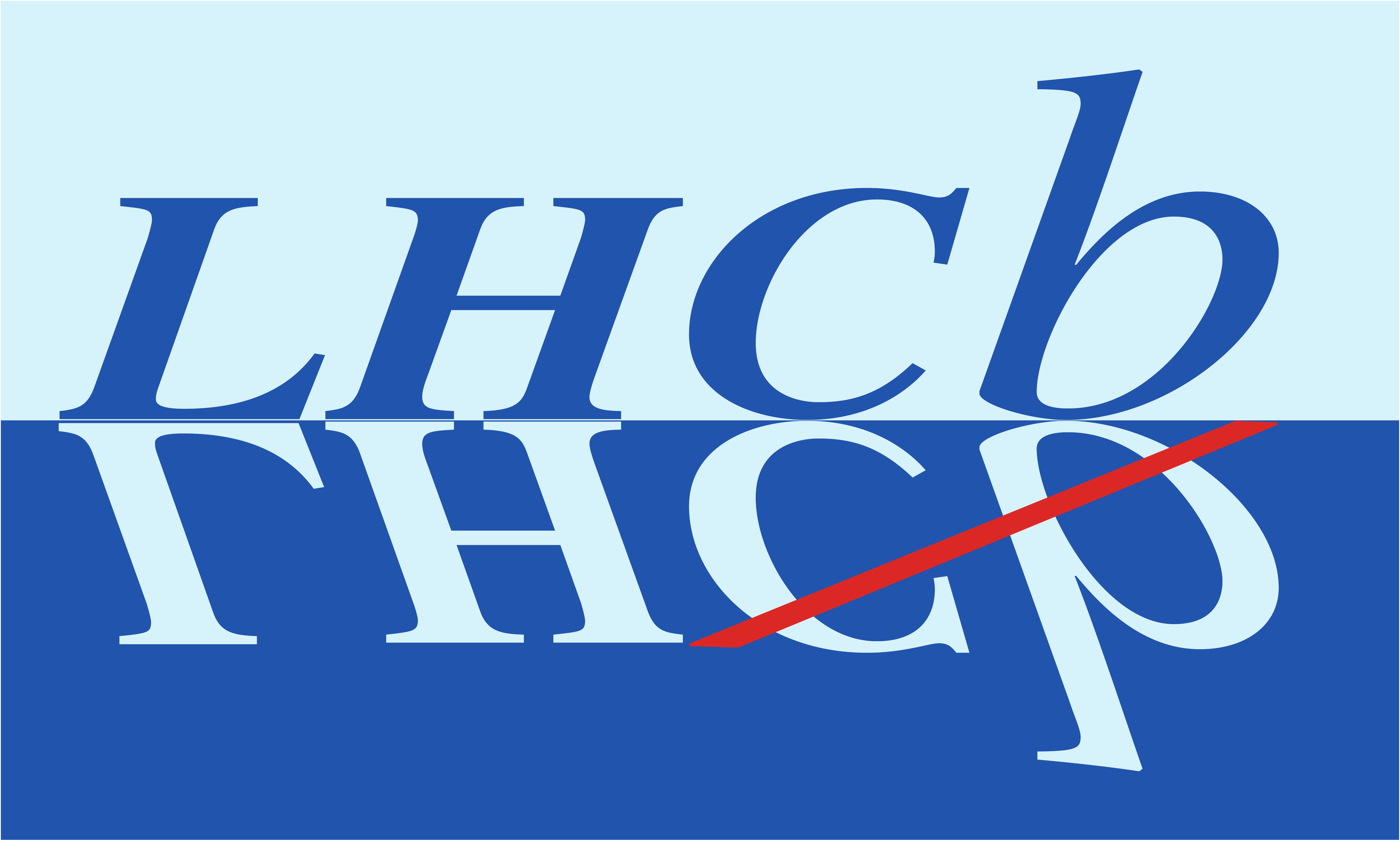}} & &}%
{\vspace*{-1.2cm}\mbox{\!\!\!\includegraphics[width=.12\textwidth]{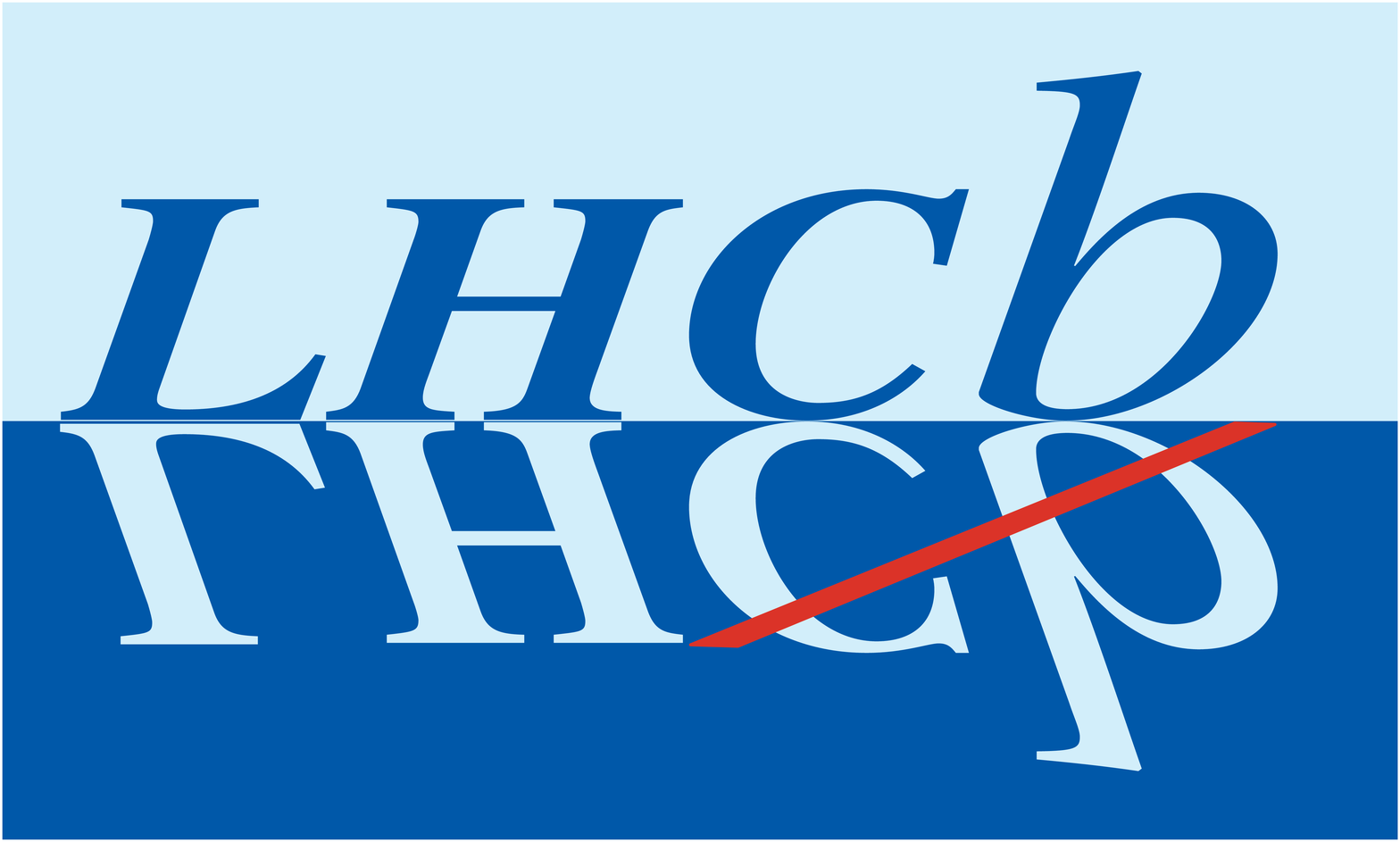}} & &}%
\\
 & & CERN-EP-2020-242 \\  
 & & LHCb-PAPER-2020-040 \\  
 & & 2 March 2021 \\ 
 & & \\
\end{tabular*}

\vspace*{4.0cm}

{\normalfont\bfseries\boldmath\huge
\begin{center}
  \papertitle 
\end{center}
}

\vspace*{2.0cm}

\begin{center}
\paperauthors\footnote{Full author list given at the end of the Letter.}
\end{center}

\vspace{\fill}

\begin{abstract}
  \noindent
  A measurement of direct \CP violation in the decay \sig ~is reported using a data sample corresponding to an integrated luminosity of 5.4\invfb, collected with the LHCb experiment at a center-of-mass energy of $\sqrt{s}=13$\tev.  The \CP asymmetry is measured to be $0.025 \pm 0.015 \pm 0.006 \pm 0.003$, where the uncertainties are statistical, systematic, and due to an external input, respectively.  This is the most precise measurement of the \CP asymmetry in the decay \sig ~and exceeds the precision on the current world average.  This direct \CP asymmetry is a key input to studies of a long-standing anomaly in \B meson decays, known as the $K\pi$-puzzle. The asymmetry is consistent with the previous measurements of this quantity, confirming and significantly enhancing the observed anomalous difference between the
direct \CP asymmetries of the $B^{0} \rightarrow K^{+} \pi^{-}$ and \sig ~decays.

\end{abstract}

\vspace*{2.0cm}

\begin{center}
  Published in
  Phys.~Rev.~Lett. 126 (2021) 091802
\end{center}

\vspace{\fill}

{\footnotesize 
\centerline{\copyright~\papercopyright. \href{\paperlicenceurl}{\paperlicence}.}}
\vspace*{2mm}

\end{titlepage}


\newpage
\setcounter{page}{2}
\mbox{~}
%
%
%
%


\renewcommand{\thefootnote}{\arabic{footnote}}
\setcounter{footnote}{0}

\cleardoublepage


\pagestyle{plain} 
\setcounter{page}{1}
\pagenumbering{arabic}


Rare decays of heavy flavored hadrons that primarily proceed through loop-level transitions are powerful probes of the effects of new physics (NP) beyond the Standard Model (SM).
The family of \decay{\B}{\kaon\pion} decays is dominated by hadronic loop  amplitudes in the SM,  but include contributions from suppressed tree-level processes, as well as electroweak loop-level processes through which NP may affect the decay~\cite{Buras:2003yc, PhysRevLett.92.101804, BURAS2004133, PhysRevD.71.057502}.   Studies of the decay $B^0\to K^+\pi^-$ at the \B-factory experiments led to the first observation of direct \CP violating asymmetries in the $B$ system~\cite{Aubert:2004qm, Chao:2004mn} resulting from the interference of two decay amplitudes where both the relative strong and weak phases are nonzero. The observed asymmetries in these modes are a result of the interference between tree- and loop-level amplitudes.  
Further studies at the \B-factory and Tevatron experiments, and at LHCb have provided measurements of the branching fractions and \CP asymmetries of the complete set of $B\to K\pi$ decays: $B^0\to K^+\pi^-$~\cite{Lees:2012kx, Duh:2012ie,Aaltonen:2014vra, LHCb-PAPER-2018-006}, $B^+\to K^+\pi^0$~\cite{Aubert:2007hh, Duh:2012ie}, $B^0\to K^0\pi^0$~\cite{Aubert:2008ad, Fujikawa:2008pk} and $B^+\to K^0\pi^+$~\cite{Aubert:2006gm, Duh:2012ie, LHCB-PAPER-2013-034}, where the inclusion of charge-conjugated processes is implied throughout this Letter, except where asymmetries are discussed. The amplitudes in the SM are expected to obey relations imposed by isospin symmetry~\cite{Buras:2003yc, PhysRevLett.92.101804, BURAS2004133, PhysRevD.71.057502, BAEK2007249, CIUCHINI2009197, BAEK200959, Gronau:2005kz, Beaudry:2017gtw, Fleischer:2017vrb}. 
However, measurements have revealed inconsistencies with this expectation.  The largest observed discrepancy is between the measured direct \CP asymmetries of the decays \decay{\Bz}{\Kp\pim} and \decay{\Bp}{\Kp\piz}.  The difference between $A_{\CP}(B^0\to K^+\pi^-)=-0.084 \pm 0.004$ and $A_{\CP}(B^+\to K^+\pi^0)=0.040 \pm 0.021$ is nonzero at 5.5 standard deviations ($\sigma$), whereas equal asymmetries are expected based on isospin arguments.  A more accurate examination of this anomaly, known as the \kaon\pion puzzle, is through the sum rule 
\begin{equation}
    A_{\CP}(\Kp\pim) + A_{\CP}(\Kz\pip)\frac{\BF(\Kz\pip)}{\BF(\Kp\pim)}\frac{\tau_0}{\tau_+} = A_{\CP}(\Kp\piz)\frac{2\BF(\Kp\piz)}{\BF(\Kp\pim)}\frac{\tau_0}{\tau_+} + A_{\CP}(\Kz\piz)\frac{2\BF(\Kz\piz)}{\BF(\Kp\pim)}
    \label{eqn:sumrule}
\end{equation}
proposed in Ref.\begin{@fileswfalse}~\cite{Gronau:2005kz}\end{@fileswfalse}, where  $A_{CP}(\kaon\pi)$ and $\BF(\kaon\pi)$ are the \CP asymmetries and the branching fractions of the $B \to \kaon\pion$ decays and $\tau_0/\tau_+$ is the ratio of the \Bz and \Bp lifetimes.  This sum rule predicts a nonzero direct asymmetry of $A_{\CP}(\Bz \to \Kz\piz) = -0.150\pm0.032$ using current world averages for the other quantities~\cite{HFLAV18}.  The current measurement of this quantity is $0.01 \pm 0.10$\cite{HFLAV18}.  The \kaon\pion puzzle has been the subject of significant theoretical attention, which includes more complete examination of the SM predictions as well as potential NP sources of the discrepancies~\cite{Buras:2003yc, PhysRevLett.92.101804, BURAS2004133, PhysRevD.71.057502, BAEK2007249, BAEK200959, Gronau:2005kz, Beaudry:2017gtw, Fleischer:2017vrb}.

This Letter presents a  measurement of direct \CP asymmetry in the decay $B^+\to K^+\pi^0$, 
\begin{equation}
    A_{\CP}=\frac{\Gamma(\Bm \to \Km \piz)-\Gamma(\Bp \to \Kp \piz)}{\Gamma(\Bm \to \Km \piz)+\Gamma(\Bp \to \Kp \piz)}
\end{equation}
 where $\Gamma(B^{\pm} \to K^{\pm} \piz)$ refers to the rate of $B^{\pm} \to K^{\pm} \piz$ decays, using data recorded with the \lhcb detector at the CERN Large Hadron  Collider.  The data sample corresponds to an integrated luminosity of 5.4\invfb collected at a center-of-mass energy of $13\tev$ between 2016 and 2018.

The \lhcb detector~\cite{LHCb-DP-2008-001, LHCB-DP-2014-002} is a single-arm forward spectrometer covering the \mbox{pseudorapidity} range $2<\eta <5$.  The detector includes a high-precision tracking system consisting of a silicon-strip vertex detector surrounding the $pp$ interaction region, a large-area silicon-strip detector located upstream of a dipole magnet with a bending power of about $4{\rm\,Tm}$, and three stations of silicon-strip detectors and straw drift tubes placed downstream of the magnet.  The tracking system provides a measurement of momentum, \ptot,  of charged particles with a relative uncertainty that varies from 0.5\% at low momentum to 1.0\% at 200\gevc.  The minimum distance of a track to a primary $pp$ collision vertex (PV), the impact parameter (IP), is measured with a resolution of $(15+29/\pt)\mum$, where \pt is the component of \ptot transverse to the beam, in \gevc.  Different types of charged hadrons are distinguished using information from two ring-imaging Cherenkov detectors (RICH).  Photons, electrons and hadrons are identified by a calorimeter system consisting of scintillating pad and preshower detectors, an electromagnetic and a hadronic calorimeter.  Charged and neutral clusters in the electromagnetic calorimeter (ECAL) are separated by extrapolating the tracks reconstructed by the tracking system to the calorimeter plane, while photons and neutral pions are distinguished by cluster shape and energy distributions.

Simulated events are used to model the effects of the detector acceptance and the imposed selection requirements.  In the simulation, $pp$ collisions are generated using \pythia~\cite{Sjostrand:2006za,*Sjostrand:2007gs} with a specific \lhcb configuration~\cite{LHCb-PROC-2010-056}.  Decays of unstable particles are described by \evtgen~\cite{Lange:2001uf}, in which final-state radiation is generated using \photos~\cite{Golonka:2005pn}.  The interaction of the generated particles with the detector and its response are implemented using the \geant toolkit~\cite{Allison:2006ve, *Agostinelli:2002hh} as described in Ref.~\cite{LHCb-PROC-2011-006}.

The decay topology $B^+ \rightarrow h^{+} \pi^0$, where $h^{+}$ is a charged hadron, presents a unique challenge in the proton-proton collision environment of the LHC.  These decays comprise a single charged track and lack a reconstructible displaced vertex, a signature typically used to identify the decays of \bquark hadrons.  The candidate selection for \sig~ candidates instead relies on identifying a charged kaon that is inconsistent with originating from any PV but consistent with originating from the $B$-meson trajectory.  That trajectory is determined by adding the momenta of the \Kp and \piz candidates, where the \piz momentum is defined as pointing from the LHCb interaction point to the coordinate of the energy deposited by the \piz candidate in the calorimeter.

The LHCb trigger system~\cite{LHCb-DP-2012-004} consists of a hardware stage, based on information from the calorimeter and muon systems, followed by a software stage, in which a full event reconstruction is applied.  Events are required to pass a hardware trigger that selects a neutral pion or photon with a high transverse energy based on energy deposits in the calorimeter.  Because of the limited ECAL position resolution a significant fraction of high \pt $\piz \to \gamma \gamma$ decays have their photons merged into a single cluster.  Only this \piz category is used in this analysis, as neutral pions with the photons resolved suffer from a large background of randomly combined clusters. Further selection relies on a dedicated software trigger developed for this analysis~\cite{LHCb-CONF-2015-001}.  A \sig~candidate is formed by adding the four-momenta of the neutral pion and a charged track identified as a kaon using information from the RICH detectors.  The charged kaon is required to have $p>12 \gevc$, $\pt>1.2 \gevc$, and a significant IP with respect to any PV.  The neutral pion is required to have $\pt>3.5 \gevc$ and the scalar sum of the \Kp and \piz $\pt$ must exceed $6.5 \gevc$.  The $B^{+}$ candidate is required to have a $\Kp\piz$ invariant mass in the range $4 \le m(\Kp\piz) \le 6.2 \gevcc$, and  $\pt>5 \gevc$.  Finally, the $B^{+}$ candidate trajectory is obtained by fixing its momentum vector to the PV with the smallest kaon IP.
The significance of the distance of closest approach between the \Kp candidate and this trajectory is denoted as DOCA-\chisq.  In order to identify \Kp candidates consistent with production via $B$-meson decay, the DOCA-\chisq is required to be small.  In the offline reconstruction, a stricter set of particle identification requirements are applied to the \Kp candidates.

Further candidate selection is based on variables characterizing how well isolated a candidate is from other tracks in the event.   Vertex-isolation variables are calculated by combining each track in the event with the \Kp candidate individually to form a two-track secondary vertex.  Three related variables are calculated: the smallest \chisq of the vertex fit between the \Kp and any other track, the smallest change in \chisq when one more track is added to that vertex, and the multiplicity of vertices having small \chisq.  The isolation of the candidate is also measured with the \pt asymmetry
\begin{equation}
\Apt = \frac{p_{{T}_{\B}} - p_{{T}_{\text{cone}}}}{p_{{T}_{\B}} + p_{{T}_{\text{cone}}}},
\end{equation}
 comparing the transverse momentum of the \Bu candidate ($p_{{T}_{\B}}$) to a scalar sum of additional charged particles nearby ($p_{{T}_{\text{cone}}}$).  Particles are considered in a cone around the reconstructed \Bu trajectory with a radius in the $\eta$--$\phi$ plane of $\Delta R \equiv \sqrt{(\Delta\phi)^2+(\Delta\eta)^2} = 1.7$, where $\Delta\phi$ is the difference in radians between the azimuthal angles of the momentum of the reconstructed \Bu candidate and the track, and $\Delta\eta$ is the difference between their pseudorapidities.  To ensure that the simulated distributions of these variables for the signal decays are consistent with data, candidate weights are generated from the ratios of these distributions between simulated \BdToKpi decays and a control sample of \BdToKpi candidates selected in the same data set.  A gradient boosted reweighter (GBR)~\cite{GBRreweight} technique is used to determine the weights, which are subsequently applied to the $B^+\to K^+\pi^0$ simulations. 
 
 The final candidate selection is performed using boosted decision tree (BDT) classifiers with the isolation variables, the  DOCA-\chisq, the smallest change in $\chi^{2}$ of the PV when including the \Kp track in the vertex fit, the \pt of the \Bu and the \Kp candidates, and the momentum of the \piz candidate as inputs.  These variables are chosen to provide discriminatory power between signal and background without biasing the $m(\Kp\piz)$ distribution.  

Two pairs of BDTs are trained and tested using data to represent background and simulated \sig~decays, corrected as described above, to represent signal.  One pair of BDTs is trained on background data with candidate invariant mass \mbox{$m(\Kp\piz)<4860 \mevcc$}, which is dominated by partially reconstructed \bquark-hadron decays.  Another pair of BDTs is trained on background data with \mbox{$m(\Kp\piz)>5700 \mevcc$}, which are primarily random \Kp \piz combinations (combinatorial background).  In each of these categories, the data sample is split randomly, a BDT classifier is trained and tested on each half, and then used to assign a score to the candidates in the other half~\cite{https://doi.org/10.1111/j.2517-6161.1974.tb00994.x, doi:10.1080/00401706.1974.10489157}.  This avoids biases due to artifacts in the training samples, while taking advantage of the full set of data available.  The optimal requirements on the classifier response variables are found for the data set by maximizing $\epsilon/\sqrt{N}$, where $\epsilon$ is the signal selection efficiency, evaluated on simulated events, and $N$ is the total number of candidates observed in a region of approximately 3 times the observed \sig~resolution around the expected \Bu mass.

Kaon candidates with $\pt>17 \gevc$ or $p>250 \gevc$ are removed from the sample after BDT selection because of insufficient coverage in the $B^{+} \rightarrow \jpsi K^{+}$ control sample described below.  They account for only $3\%$ of the candidates after final selection.

\begin{figure}[tb]
	\centering
	\includegraphics[width=1.0\linewidth]{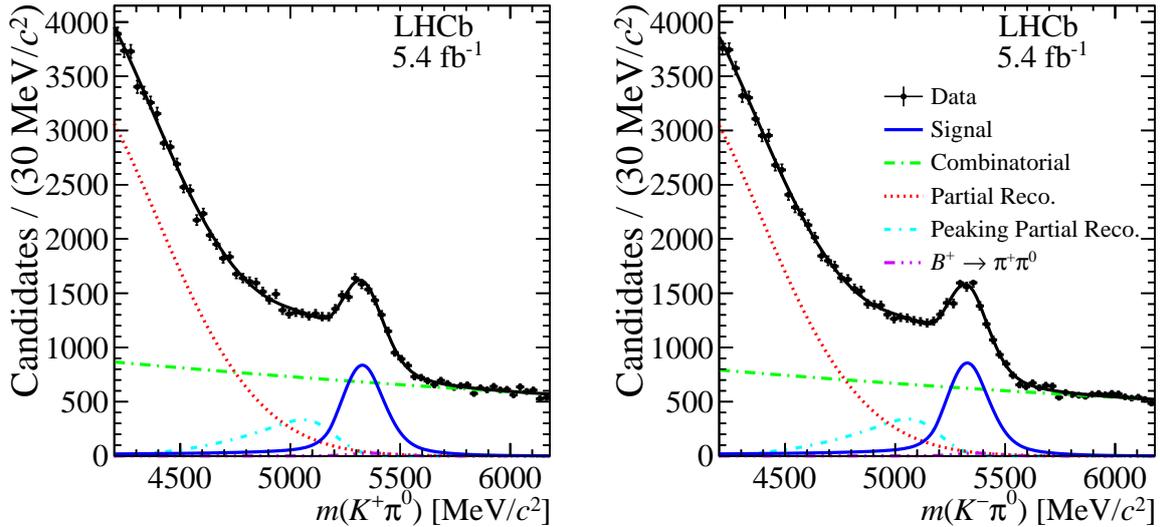}	
	\caption{Invariant-mass distribution of the selected candidates with fit projections overlayed. The data set is divided by the charge of the $B$ meson, with $B^+ \to K^+\pi^0$ shown on the left and $B^- \to K^-\pi^0$ on the right.}
	\label{fig:ACPfit}
\end{figure}

The $m(\Kp\piz)$ distribution of the selected \sig~candidates, separated by the charge of the $B$ meson, is shown in Fig.~\ref{fig:ACPfit} along with the results of a fit to the data.  In the fit, the signal is modeled by the sum of a Crystal Ball function~\cite{Skwarnicki:1986xj} and a Gaussian function with an exponential tail describing the high-mass region.  The Crystal Ball and the Gaussian functions share a common mean and width that varies freely in the fit, and their tail shape parameters are fixed from simulation. 
Combinatorial background is modeled by an exponential function, with the exponent parameter allowed to vary freely in the fit. The tail of a Gaussian function is used to model the partially reconstructed background in the low-mass region, with mean and width allowed to vary freely in the fit.  The rate of $\pip \to \Kp$ misidentification is measured in $D^{0} \rightarrow K^{-} \pi^{+}$ decays as a function of pion momentum and pseudorapidity, with the same particle identification requirements as signal events~\cite{LHCb-PUB-2016-021}.  The contribution of the misidentified $B^{+} \rightarrow \pi^{+}\pi^{0}$ background is inferred from its branching fraction~\cite{PDG2020} and the misidentification rate to be $2.4\%$ of the \sig~yield.  The $B^{+} \rightarrow \pi^{+}\pi^{0}$ background component is modeled by a Gaussian with mean and resolution fixed to values determined from simulated events and a yield fixed to the expectation.  There is assumed to be no asymmetry in this background.

An additional class of background candidates arises from decays such as  \mbox{$B^{+} \rightarrow (K^{*+} \rightarrow K^{+} \pi^{0}) \pi^{0}$}, $B^{0} \rightarrow (K^{*0} \rightarrow K^{+} \pi^{-}) \pi^{0}$, and $B^{0} \rightarrow K^{+}(\rho^{-} \rightarrow \pi^{-} \pi^{0})$ where a pion from the $K^{*}$ or $\rho^{-}$ decay is not reconstructed.  The polarization of the $K^{*}$ or $\rho^{-}$ meson results in a double peaked $m(\Kp\piz)$ distribution, where the higher-mass peak is close to the expected $B$ mass. This type of background is modeled with a parabolic function convolved with a Gaussian resolution function following the method described in Ref.~\cite{LHCb-PAPER-2019-044}.  The width of the resolution function is fixed to that of the signal resolution and the end points are fixed to the kinematic end points, allowing for a shift between the fitted and the known \Bu masses\cite{PDG2020}.  The lower-mass peak contributes below the $m(\Kp\piz)$ range considered, and so its relative height is fixed to a value determined from simulation.

Other background sources include: $B^{+} \rightarrow (K^{*+} \rightarrow K^{+} \pi^{0}) \gamma$ decays where the $\gamma$ is misidentified as a $\pi^{0}$; $B^{+}\rightarrow (f^{0}(980) \rightarrow \pi^{0} \pi^{0}) K^{+}$ decays where one $\pi^{0}$ is not reconstructed; and $B^{0} \rightarrow (\bar{D}^{0} \rightarrow K^{+} \pi^{-}) \pi^{0}$, $B^{0}\rightarrow(K^{*}_{0}(1430) \rightarrow K^{+} \pi^{-}) \piz$  and  $B^{0}_{s} \rightarrow K^{+} \pi^{0} \pi^{-}$ decays where the $\pi^{-}$ is not reconstructed.  Simulation studies have shown that these background contributions either have $m(\Kp\piz)$ distributions indistinguishable from the partially reconstructed samples described by a Gaussian tail, or in the case of $B^{0}_{s} \rightarrow K^{+} \pi^{0} \pi^{-}$ has a branching fraction too small to give an observable contribution.

The data are fitted separately in four categories.  In order to reduce uncertainties due to non-uniformity of the detector, candidates are separated according to whether the LHCb dipole magnetic field is aligned vertically upward (Magnet Up) or downward (Magnet Down) in the experiment.  Candidates are further separated by $B$-meson charge in order to measure the \CP asymmetry.  The yield and asymmetry of each fit component are allowed to vary freely while the shape parameters are the same for the $B^{+}$ and $B^{-}$ candidates.  The total yield of $B^{\pm} \to K^{\pm}\piz$~decays is measured to be $8310 \pm 255$ in the Magnet Up data set and $8373 \pm 253$ in the Magnet Down data set.  The raw asymmetry, $A_{\textrm{raw}}$, between the $B^-$ and $B^+$ signal yields is found to be $0.019 \pm 0.021$ for Magnet Down and $0.005 \pm 0.022$ for Magnet Up.  The results are consistent when candidates are separated by data-taking year as well as when shape parameters are allowed to vary independently for all four data categories.

As the measured asymmetry receives contributions from a number of nuisance asymmetries, the \CP asymmetry can be expressed as
\begin{equation}
A_{\CP}(\sig) = A_{\textrm{raw}}(\sig)-A^{B}_{\textrm{prod.}}-A^{K}_{\textrm{det.}},
\label{eq1}
\end{equation}
where $A^{B}_{\textrm{prod.}}$ is the production asymmetry of $B^{\pm}$ mesons and $A^{K}_{\textrm{det.}}$ is the combined asymmetry in detection, triggering, and reconstruction of $K^{\pm}$ mesons.  These effects must be corrected for in order to extract $A_{\CP}$ from $A_{\textrm{raw}}$.  The combined effect of the nuisance asymmetries is measured with a control sample of $B^{+} \rightarrow (J/\psi \rightarrow \mu^{+} \mu^{-}) K^{+}$ decays, using the same data sample as the signal channel. 

In the hardware trigger, events with a $B^{+} \rightarrow (\jpsi \rightarrow \mu^{+} \mu^{-}) K^{+}$ decay are required to trigger on particles other than the kaon, in order to avoid introducing additional trigger asymmetries.  At the software stage the event must trigger on the kaon in the same manner as signal events.  The offline selection requires that the $B$-meson lifetime be greater than 0.1\ps and that the kaon and muons have a significant IP with respect to all PVs.  Additional requirements on the momentum of the kaon and $B$ candidates as well as kaon particle identification are imposed to match the signal selection.  The momentum distributions of the $B^+$ and $K^{+}$ candidates are weighted to match those of the signal candidates using the GBR technique~\cite{GBRreweight}, as the production and detection asymmetries may depend on kinematics of the decay.  

The raw asymmetry in the $B^{+} \rightarrow \jpsi K^{+}$ signal yields is determined via an unbinned maximum-likelihood fit in which the invariant-mass distribution of the $B^{+} \rightarrow \jpsi K^{+}$ candidates is modeled by the sum of two Gaussian functions sharing a common mean, while the combinatorial background is modeled by an exponential distribution. The total yield of $B^{+} \rightarrow \jpsi K^{+}$ decays is measured to be $372874 \pm 776$ for Magnet Down and $306821 \pm 699$ for Magnet Up data samples with a purity of approximately 99\%. The raw asymmetry is found to be $-0.009 \pm 0.002$ for Magnet Up, and $-0.012 \pm 0.002$ for Magnet Down samples.  The \CP asymmetry for the decay $\B^{+} \rightarrow (\jpsi \rightarrow \mu^{+} \mu^{-}) K^{+}$ is taken to be \mbox{$A_{\CP}(B^{+} \to \jpsi K^{+})=0.002 \pm 0.003$} from Ref.~\cite{PDG2020}. After subtracting $A_{\CP}$, the remaining asymmetry is attributed to the combination of production, detection, reconstruction, and triggering effects, which can then be determined from
\begin{equation}
A^{B}_{\textrm{prod.}}+A^{K}_{\textrm{det.}} = A_{\textrm{raw}}(B^{+} \to \jpsi K^{+}) - A_{\CP}(B^{+} \to \jpsi K^{+}).
\label{eq2}
\end{equation}
This estimate of the nuisance asymmetry is then used in Eq.~\ref{eq1} to determine $A_{\CP}(\Bp \to \Kp \piz)$.This is done separately for the Magnet Up and Magnet Down data.  By averaging the Magnet Up and Magnet Down results, the direct \CP asymmetry is determined to be $A_{\CP}(\sig)=0.025 \pm 0.015$, where the uncertainty is statistical only.

\begin{table}[t]
	\caption{Systematic uncertainties on $A_{\CP}(\sig)$.}
	\centering
	\begin{tabular}{lc}
	\toprule
		Systematic & Value ($\times 10^{-3}$)\\
		Signal modeling shape  & $4.3$\\
		Combinatorial background shape & $1.3$\\
		Partial reco. background shape& $1.3$\\
		Peaking partial reco. background shape & $1.2$\\
     	Peaking partial reco. background offset & $1.3$\\
		Peaking partial reco. background resolution & $1.4$\\
		$\Bp \rightarrow \pip \piz$ yield & $1.3$\\
		$\Bp \rightarrow \pip \piz$ \CP asymmetry& $1.5$\\
		Multiple candidates & $1.3$\\
		Production/detection asymmetry stat. & $2.1$  \\
		Production/detection asymmetry weights & $0.5$  \\
        \midrule
		Sum in quadrature & $6.1$ \\
		\bottomrule
	\end{tabular}
	\label{tab:systs}
\end{table}

To assess the systematic uncertainty due to mismodeling of the signal and background line shapes, pseudoexperiments are generated for variations of the $m(\Kp\piz)$ fit model.  The leading source of systematic uncertainty is from modeling the signal component in the fit. This uncertainty is assessed by replacing the default model with a single Gaussian distribution.  Systematic uncertainties are assessed for numerous fit variations: replacing the exponential distribution for the combinatorial background with a linear function, individually replacing each low-mass background model with an Argus function, allowing the position and resolution of the peaking low-mass background to vary freely and independently of the signal distribution, and varying the yield and asymmetry of $B^{+} \rightarrow \pi^{+}\piz$  background.  Pseudoexperiments are also generated to assess the systematic uncertainty due to including events with multiple candidates in the base analysis.

The statistical uncertainty on the determination of the raw $B^{+} \rightarrow \jpsi K^{+}$ asymmetry is also considered as a systematic uncertainty, and is the subdominant source of systematic uncertainty.  Additionally, the difference between the nuisance asymmetries with and without applying the GBR weights is taken to be a systematic uncertainty.  The estimated values for all systematic uncertainties are shown in Table~\ref{tab:systs}, where the common value of $0.0013$ is from the statistical uncertainty of the pseudoexperiments generated. The $A_{\CP}(B^{+} \to \jpsi K^{+})$ precision of 0.003 is considered separately as an external-input uncertainty.

In conclusion, the direct \CP asymmetry of the decay \sig\ has been measured with the LHCb detector using a data sample corresponding to a luminosity of 5.4 \invfb.   It is found to be 
\begin{equation}\nonumber
A_{\CP}(\sig) = 0.025 \pm 0.015 \pm 0.006 \pm 0.003, 
    \label{eqn:result}
\end{equation}
where the first uncertainty is statistical, the second is systematic and the third due to external inputs, exceeding the precision of the current world average~\cite{HFLAV18}.  This result is consistent with the world average and consistent with zero at approximately $1.5\,\sigma$.  The \CP asymmetry difference,  $\Delta A_{\CP}(\kaon\pion) \equiv A_{\CP}(\sig{}) - A_{\CP}(\BdToKpi)$, is found to be $0.108 \pm 0.017$, where $A_{\CP}(\BdToKpi)$ is taken from Ref.~\cite{HFLAV18}\footnote{The world average includes the LHCb measurement, and a small correlation between the LHCb measurements of $A_{\CP}(\BdToKpi)$ and $A_{\CP}(\sig)$ due to the charged kaon detection asymmetry has been neglected}.  
Including the result presented in this Letter, the new world average of $A_{\CP}(\sig{})$ is found to be  $0.031 \pm 0.013$. 
This corresponds to $\Delta A_{\CP}(\kaon\pion)=0.115 \pm 0.014$, which is nonzero with a significance of more than 8 standard deviations, substantially enhanced over the results prior to this measurement.  The updated sum rule prediction for $A_{\CP}(\Bz \to \Kz\piz)$, shown in Eq.~\ref{eqn:sumrule}, is found to be $-0.138\pm0.025$, departing from zero with a significance of approximately $5.5\,\sigma$.

\section*{Acknowledgements}
%
%
\noindent We express our gratitude to our colleagues in the CERN
accelerator departments for the excellent performance of the LHC. We
thank the technical and administrative staff at the LHCb
institutes.
We acknowledge support from CERN and from the national agencies:
CAPES, CNPq, FAPERJ and FINEP (Brazil); 
MOST and NSFC (China); 
CNRS/IN2P3 (France); 
BMBF, DFG and MPG (Germany); 
INFN (Italy); 
NWO (Netherlands); 
MNiSW and NCN (Poland); 
MEN/IFA (Romania); 
MSHE (Russia); 
MICINN (Spain); 
SNSF and SER (Switzerland); 
NASU (Ukraine); 
STFC (United Kingdom); 
DOE NP and NSF (USA).
We acknowledge the computing resources that are provided by CERN, IN2P3
(France), KIT and DESY (Germany), INFN (Italy), SURF (Netherlands),
PIC (Spain), GridPP (United Kingdom), RRCKI and Yandex
LLC (Russia), CSCS (Switzerland), IFIN-HH (Romania), CBPF (Brazil),
PL-GRID (Poland) and OSC (USA).
We are indebted to the communities behind the multiple open-source
software packages on which we depend.
Individual groups or members have received support from
AvH Foundation (Germany);
EPLANET, Marie Sk\l{}odowska-Curie Actions and ERC (European Union);
A*MIDEX, ANR, Labex P2IO and OCEVU, and R\'{e}gion Auvergne-Rh\^{o}ne-Alpes (France);
Key Research Program of Frontier Sciences of CAS, CAS PIFI, CAS CCEPP, 
Fundamental Research Funds for Central Universities, 
and Sci. \& Tech. Program of Guangzhou (China);
RFBR, RSF and Yandex LLC (Russia);
GVA, XuntaGal and GENCAT (Spain);
the Royal Society
and the Leverhulme Trust (United Kingdom).




\addcontentsline{toc}{section}{References}
\bibliographystyle{LHCb}
\bibliography{main,standard,LHCb-PAPER,LHCb-CONF,LHCb-DP,LHCb-TDR}

\newpage
\centerline
{\large\bf LHCb collaboration}
\begin
{flushleft}
\small
R.~Aaij$^{32}$,
C.~Abell{\'a}n~Beteta$^{50}$,
T.~Ackernley$^{60}$,
B.~Adeva$^{46}$,
M.~Adinolfi$^{54}$,
H.~Afsharnia$^{9}$,
C.A.~Aidala$^{85}$,
S.~Aiola$^{25}$,
Z.~Ajaltouni$^{9}$,
S.~Akar$^{65}$
% ...

{\footnotesize \it

% ...
$^{9}$Universit{\'e} Clermont Auvergne, CNRS/IN2P3, LPC, Clermont-Ferrand, France\\
% ...
$^{25}$INFN Sezione di Milano, Milano, Italy\\
% ...
$^{32}$Nikhef National Institute for Subatomic Physics, Amsterdam, Netherlands\\
% ...
$^{46}$Instituto Galego de F{\'\i}sica de Altas Enerx{\'\i}as (IGFAE), Universidade de Santiago de Compostela, Santiago de Compostela, Spain\\
% ...
$^{50}$Physik-Institut, Universit{\"a}t Z{\"u}rich, Z{\"u}rich, Switzerland\\
% ...
$^{54}$H.H. Wills Physics Laboratory, University of Bristol, Bristol, United Kingdom\\
% ...
$^{60}$Oliver Lodge Laboratory, University of Liverpool, Liverpool, United Kingdom\\
% ...
$^{65}$University of Cincinnati, Cincinnati, OH, United States\\
% ...
$^{68}$Syracuse University, Syracuse, NY, United States\\
% ...
$^{85}$University of Michigan, Ann Arbor, United States, associated to $^{68}$\\
\bigskip
% ...
\medskip
}
\end{flushleft}

\end{document}